\documentclass[a4paper,11pt]{article}
\pdfoutput=1 
\usepackage[utf8]{inputenc}
\usepackage{jheppub} 
\usepackage{slashed}                     
\usepackage[T1]{fontenc} 
\graphicspath{{img/}} 

\title{\boldmath Total and Symmetry resolved Entanglement spectra in some Fermionic CFTs from the BCFT approach}

 \author{Himanshu Gaur}
 \affiliation{Department of Physics, Indian Institute of Technology Bombay, Powai, Mumbai, Maharashtra 400076 India}

\emailAdd{194123018@iitb.ac.in}

\abstract{In this work, we study the universal total and symmetry-resolved entanglement spectra for a single interval of some $2$d Fermionic CFTs using the Boundary Conformal Field theory (BCFT) approach. In this approach, the partition of Hilbert space is achieved by cutting out discs around the entangling boundary points and imposing boundary conditions preserving the extended symmetry under scrutiny. The reduced density moments are then related to the BCFT partition functions and are also found to be diagonal in the symmetry charge sectors. In particular, we first study the entanglement spectra of massless Dirac fermion and modular invariant $Z_2$-gauged Dirac fermion by considering the boundary conditions preserving either the axial or the vector $U(1)$ symmetry. The total entanglement spectra of the modular invariant $Z_2$-gauged Dirac fermion are shown to match with the compact boson result at the compactification radius where the Bose-Fermi duality holds, while for the massless Dirac fermion, it is found that the boundary entropy term doesn't match with the self-dual compact boson. The symmetry-resolved entanglement is found to be the same in all cases, except for the charge spectrum which is dependent on both the symmetry and the theory. We also study the entanglement spectra of $N$ massless Dirac fermions by considering boundary conditions preserving different chiral $U(1)^N$ symmetries. Entanglement spectra are studied for $U(1)^M$ subgroups, where $M\leq N$, by imposing boundary conditions preserving different chiral symmetries. The total entanglement spectra are found to be sensitive to the representations of the $U(1)^M$ symmetry in the boundary theory among other behaviours at $O(1)$. Similar results are also found for the Symmetry resolved entanglement entropies. The characteristic $\log\log\left(\ell/\epsilon\right)$ term of the $U(1)$ symmetry is found to be proportional to $M$ in the symmetry-resolved entanglement spectra.}

\makeatletter
\gdef\@fpheader{}
\makeatother
\begin{document} 
\maketitle
\flushbottom

\section{Introduction} \label{introduction}
The study of the entanglement phenomenon in quantum theories has proven to be one of the most rewarding programs in recent times. It has provided new insights in many fields of physics, perhaps most notable among these are quantum information theory \cite{nielsen2010quantum}, gauge-gravity duality \cite{ryu2006holographic, ryu2006aspects, nishioka2009holographic}, Black hole physics \cite{solodukhin2011entanglement}, and quantum many-body systems \cite{amico2008entanglement, vidal2003entanglement}. In particular for low dimensional critical systems, the entanglement spectrum is sensitive to the corresponding CFT data \cite{calabrese2004entanglement, calabrese2009entanglement2, coser2014renyi}. 

In bipartite entanglement studies, one considers a bipartition of the system into subsystem A and its complement B in the system. This partition is such that the total Hilbert space decomposes into the Hilbert spaces of subsystem $A$, and $B$, mathematically $\mathcal{H}=\mathcal{H}_A\otimes\mathcal{H}_B$. An effective class of entanglement measures are R\'enyi entropies, sometimes also called entanglement entropies, they are defined as
\begin{equation}
S_n=\frac{1}{1-n}\log\left(\rho_A^{n}\right),
\end{equation}
where $\rho_A$ is the reduced density matrix. It is given by the trace of the density matrix $\rho$ of the system over degrees of freedom in the subsystem $B$, mathematically denoted by $\rho_A=\mathrm{Tr}_B\rho$. We mention here that these measures of entanglement fail when we are dealing with mixed-state density matrices since they fail to distinguish the quantum correlations from the classical ones. In the present work, we will only be dealing with the pure states.

For the conformal field theories in the ground state, the R\'enyi entropies for a single interval are proportional to the central charge $c$ and scale as the logarithm of the interval length to the leading order \cite{calabrese2004entanglement, holzhey1994geometric}. When the subsystem $A$ is composed of multiple disjoint intervals the R\'enyi entropies are also sensitive to the local conformal operator content as well \cite{calabrese2009entanglement2, coser2014renyi, furukawa2009mutual}.

In such entanglement studies, especially when the theory is in the continuum, there arises a subtlety while defining the partition. In quantum field theories, the decomposition of the Hilbert space $\mathcal{H}=\mathcal{H}_A\otimes\mathcal{H}_B$ cannot be defined by a sharp cutting operation \cite{ohmori2015physics, witten2018aps}. One solution to this problem is to cut out small discs around the boundaries and impose appropriate boundary conditions on the discs \cite{ohmori2015physics, cardy2016entanglement}. This is particularly interesting in conformal field theories, as it leads to the study of new universal contributions to the entanglement spectrum when the imposed boundary conditions are conformally invariant. We will discuss this procedure further, relevant to our present program, in Section \ref{section2}. Such contributions have been studied in the quantum many-body systems in \cite{ohmori2015physics, cardy2010unusual, hsu2009universal, cardy2014universal}.  

Symmetry has been the cornerstone in the study of physical systems. Its utility in improving our understanding of the quantum phenomenon is unparalleled. It was noticed in \cite{goldstein2018symmetry} that when the system possesses a global internal symmetry, its bipartite entanglement may be resolved into the local charge sectors of the subsystem $A$ under certain conditions. This study allows us to study the resolution of entanglement into the symmetry sectors thus improving our understanding of quantum correlations in such systems. Such studies have been termed symmetry-resolved entanglement. Some recent experimental results \cite{lukin2019probing, neven2021symmetry} in this direction have generated considerable attention towards such studies. The symmetry resolved entanglement entropies, largely for $U(1)$ symmetries, has been widely studied in quantum many-body and critical systems \cite{goldstein2018symmetry, xavier2018equipartition, turkeshi2020entanglement, bonsignori2019symmetry,  fraenkel2020symmetry,ares2022symmetry, jones2022symmetry,2023,barghathi2019operationally, barghathi2018renyi,ghasemi2023universal, murciano2020entanglement, murciano2020symmetry, murciano2020symmetry1, ares2022multi, gaur2024multi, foligno2023entanglement, capizzi2022entanglement, horvath2021u, capizzi2022renyi, capizzi2022renyi2, capizzi2023full, parez2021exact, parez2021quasiparticle, estienne2021finite, di2023boundary, kusuki2023symmetry, calabrese2021symmetry, pirmoradian2023symmetry}.  The symmetry resolution of negativity \cite{cornfeld2018imbalance,murciano2021symmetry, gaur2023charge, chen2022charged, chen2022dynamics,chen2023dynamics, feldman2019dynamics, parez2022dynamics, berthiere2023reflected, bruno2023symmetry}, relative entropy \cite{capizzi2021symmetry, chen2021symmetry} and operator entanglement \cite{rath2023entanglement, murciano2023more} have also received some attention in similar settings. Symmetry-resolved entanglement and related quantities have also been studied in the context of gravity \cite{zhao2021symmetry, weisenberger2021symmetry, zhao2022charged, belin2013holographic, milekhin2021charge, gaur2023symmetry}. Recently Entanglement asymmetry, a measure used to probe the degree of broken symmetry, has also been studied in quantum many-body systems \cite{ares2023entanglement, ares2023lack, ferro2023non, murciano2023entanglement, capizzi2023universal, capizzi2023entanglement, chen2023entanglement}. In the context of the Boundary conformal field theory (BCFT) approach, mentioned earlier, the symmetry-resolved entanglement has been studied in \cite{di2023boundary, kusuki2023symmetry, PhysRevLett.131.151601}.

In the present work, we will be interested in total and symmetry-resolved entanglement spectrum of some $2$d Fermion conformal theories in the context of Hilbert space decomposition mentioned above. In \cite{di2023boundary}, it was noticed that when the boundary conditions imposed on the discs preserve the symmetry under scrutiny, it is possible to study all universal contributions to the symmetry-resolved entanglement entropies. The specific case of compact boson was considered in the Reference above. We will consider massless Dirac fermions and $Z_2$-gauged Dirac fermions for these studies first. Since the $Z_2$-gauged Dirac fermions considered are modular invariant, we will refer to them as modular invariant Dirac fermions in this work, see Section \ref{section3.2}. Both the theories possess $U(1)$ symmetries, with respect to which we study the symmetry resolution. We also compare the results with the boson case motivated by the Bose-Fermi duality in 2d CFT. We then consider $N$ massless Dirac fermions and consider the $U(1)^M$ chiral symmetries studied in \cite{smith2020boundary}, where $M\leq N$, in the same context. These symmetries are subgroups of the total symmetry group $SO(2N)\times SO(2N)$, thus allowing to study the symmetry resolution of entanglement for smaller subgroups of the symmetry group.

The organisation of the work is as follows. In Section \ref{section2}, we review symmetry resolved entanglement and the Boundary Conformal Field Theory (BCFT) approach to study the total and symmetry resolved entanglement spectrum for R\'enyi entropies. In Section \ref{section3}, we study both the R\'enyi entropies of the massless Dirac fermion and modular invariant Dirac fermions. In Section \ref{section4}, we study the same R\'enyi entropies of $N$ massless Dirac fermions. The boundary conditions preserving $U(1)^N$ chiral symmetries are considered to study the entanglement spectrum and the symmetry resolved entanglement spectrum $U(1)^M$ subgroups, where $M\leq N$. We conclude our work in Section \ref{section5}. Finally, we also have two Appendices \ref{A}, and \ref{B}, containing necessary calculations and conventions used in the main text.    
\section{BCFT and Symmetry resolution of Entanglement} \label{section2}
We will study the universal total and symmetry resolved entanglement spectrum for a single interval $A$ as shown in Figure \ref{fig:i}. Here we first discuss the symmetry resolution of entanglement entropies for theories with a global internal symmetry \cite{goldstein2018symmetry}. We then discuss the BCFT approach to study the same when the theory under consideration is a CFT. The BCFT approach bypasses the calculation of charged moments usually involved in such studies and also allows us to find the complete universal spectrum of entanglement.
\subsection{Symmetry resolved Entanglement}
We consider a theory with global internal symmetry and take the bipartition of the system into subsystem $A$, and its complement $B$. We assume that the theory is in a state $\rho$, such that $[\rho,\hat{Q}]=0$, where $\hat{Q}$ is the charge corresponding to the global internal symmetry. If the charge is additive in subsystems $A$, and $B$, then we have $[\rho_A,\hat{Q}_A]=0$. This implies that $\rho_A$ is block diagonal in local charge sectors $q$ in subsystem A. This allows us to study entanglement in the local charge sectors. Such studies have been termed symmetry-resolved entanglement. Since we are only interested in pure states in this work, we restrict ourselves to pure density matrices $\rho$ from here onwards. To study entanglement in local charge sectors we define \cite{bonsignori2019symmetry}
\begin{equation} \label{eq2.1}
\rho_{A,q}=\frac{\Pi_q\rho_A\Pi_q}{\mathrm{Tr}\left(\Pi_q\rho_A\right)},
\end{equation}
where $\Pi_q$ is a unitary projector onto the local charge sector $q$. Note that the trace of $\rho_{A,q}$ is unity. The quantity $\mathrm{Tr}\left(\Pi_q\rho_A\right)$ is just the probability of obtaining the charge $q$ upon measurement, let us denote this quantity by $p_q$ from here on. The symmetry resolved R\'enyi entropies $S_{n,q}$ are given by
\begin{equation} \label{eq2.2}
S_{n,q}=\frac{1}{1-n}\log\left[\mathrm{Tr}\rho_{A,q}^n\right].
\end{equation}
However, the $\Pi_q$ are not always readily constructed. To overcome this, for instance, charge moments are introduced when the global symmetry under consideration is $U(1)$. As will be discussed in the subsequent subsection, the BCFT approach already gives a decomposition of moments of reduced matrix into the local charge sectors thus allowing us to study the symmetry-resolved entanglement directly. Now from eq.\eqref{eq2.2}, we may write the R\'enyi entropy $S_n$ as
\begin{equation} \label{eq2.3}
S_n=\frac{1}{1-n}\log\left[\sum_q p_q^n \mathrm{Tr}\rho_{A,q}^n\right].
\end{equation}
Then the entanglement entropy $S_1$ has two contributions
\begin{equation} \label{eq2.4}
\begin{split}
S_1&=-\sum_q p_q\log p_q-\sum_q p_q \mathrm{Tr}\left[\rho_{A,q}\log\rho_{A,q}\right]\\
&=S_{1,f}+S_{1,c},
\end{split}
\end{equation}
where $S_{1,c}$, known as the configuration entropy, is just the sum of entropies in each local charge sector weighted by the respective probabilities $p_q$ \cite{lukin2019probing}. The contribution $S_{1,f}$, known as fluctuation entropy, is the entropy due to the probability distribution of the local charge $q$ \cite{lukin2019probing}. A similar decomposition for all R\'enyi entropies is possible when we have the equipartition of R\'enyi entropies, i.e. eq.\eqref{eq2.2} is independent of $q$. In such cases we have the following decomposition of R\'enyi entropy \cite{di2023boundary}
\begin{equation} \label{eq2.5}
\begin{split}
S_n&=\frac{1}{1-n}\log\left[ \sum _q p_q^n\right]+\frac{1}{1-n}\sum_q p_q \log \rho_{A,q}^n\\
&=S_{n,f}+S_{n,c}.
\end{split}
\end{equation}
The entropies $S_{n,f}$ and $S_{n,c}$ here have the same interpretation as above. As we'll see in the next subsection, the BCFT approach also makes it convenient to study the fluctuation entropy as well.
\subsection{BCFT approach} 
To study entanglement for a bipartition, one begins with the decomposition of the Hilbert space $\mathcal{H}$ in $\mathcal{H}_A\otimes\mathcal{H}_B$ (or at least the decomposition of the Hilbert space sector in which the state of the system lies). This is not a straightforward exercise in field theories for spatial entanglement, since the Hilbert space $\mathcal{H}$ is built by acting the smeared fields on the vacuum state $\left|\Omega\right.\rangle$. This is done to ensure that the states thus created have finite norms. So a simple sharp cutting of space into spatial regions $A$, and $B$ doesn't give a locally well-defined decomposition of the Hilbert space $\mathcal{H}$ \cite{ohmori2015physics, witten2018aps}. As discussed in \cite{cardy2016entanglement}, a well-defined procedure is to consider the projection of states in the Hilbert space $\mathcal{H}$ onto a common eigenstate of a complete set of commuting local operators in some open set containing the boundary between the spatial regions $A$, and $B$. In path integral language this is equivalent to imposing some appropriate boundary conditions on the boundary of the open set and cutting out the open set from the geometry. This cut in space is characterised by a length parameter $\epsilon$. The parameter $\epsilon$ acts to regularise the entanglement measures.

Let us now restrict our attention to CFTs. In CFTs, we may choose conformally invariant boundary conditions. Using these boundary conditions we obtain the universal contributions to the entanglement spectra \cite{cardy2016entanglement}. It is also possible to obtain the non-universal contributions to entanglement spectra by considering appropriate boundary conditions \cite{ohmori2015physics, cardy2016entanglement}. However, we will restrict our attention to boundary conditions which respect the conformal invariance in the present work. In what follows we assume without the loss of generality that these cuts are circular discs of radius $\epsilon$ centred at the boundary between $A$, and $B$ \cite{ohmori2015physics}.
\begin{figure}
\centering 
\includegraphics[width=0.6\textwidth]{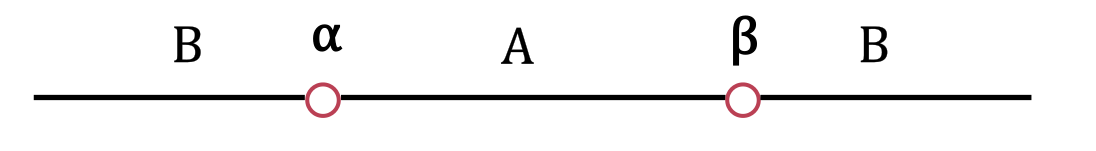}
\caption{\label{fig:i} Spatial cuts introduced to partition the space into spatial region $A$, and $B$. Boundary conditions $\alpha$ and $\beta$ are introduced on the boundary of these cuts.}
\end{figure}

\begin{figure}
\centering 
\includegraphics[width=0.9\textwidth]{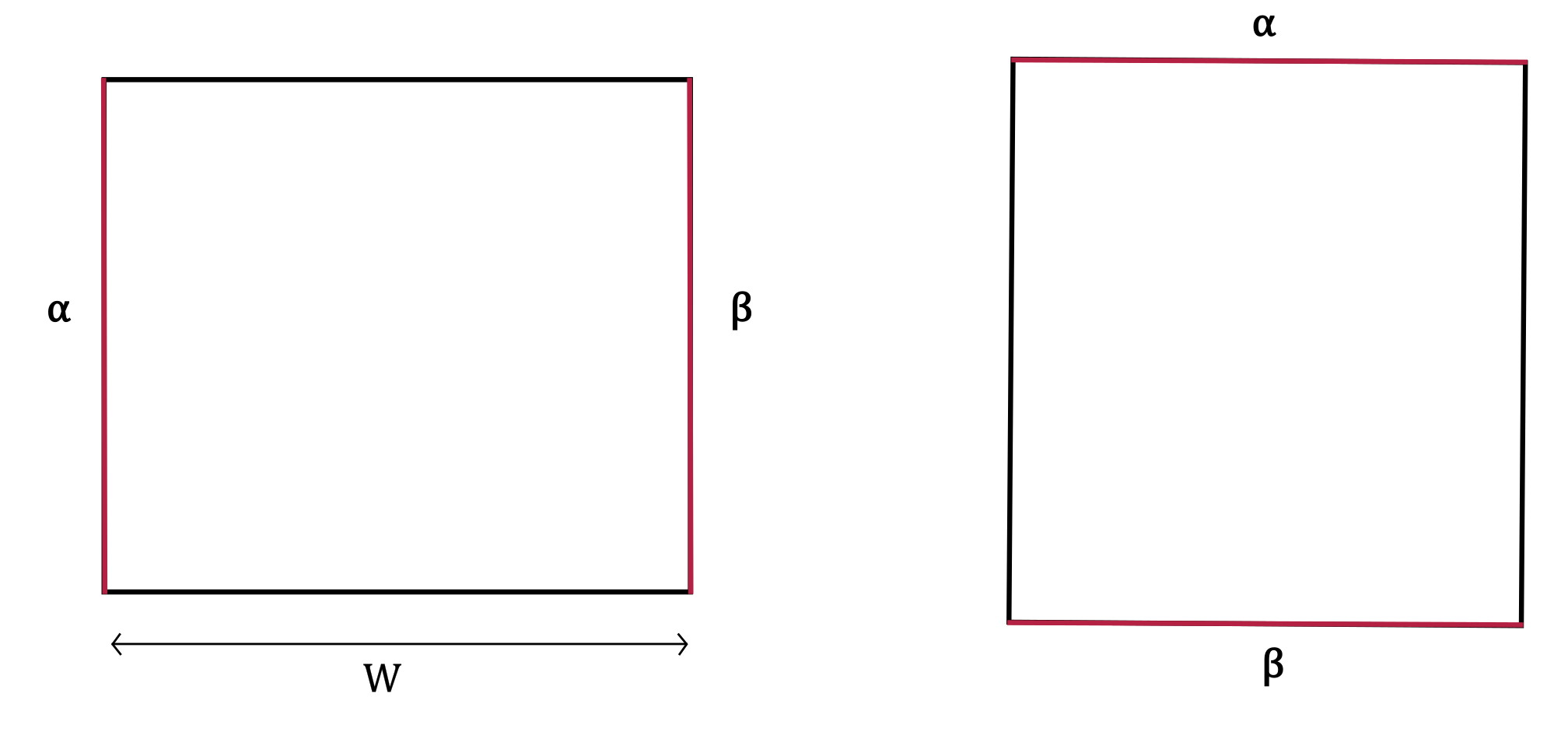}
\caption{\label{fig:ii} Annulus with the boundary conditions $\alpha$ and $\beta$ are introduced in the open string (on left) and closed string (on right) channel.}
\end{figure}

We recall that evaluation of the moments of reduced density matrix $\rho_A$ involves taking $n$-copies of the geometry in Figure \ref{fig:i}, pasting them together and computing the partition function on the resulting geometry. The associated geometry in our present program is an annulus. Consider the following conformal transformation
\begin{equation} \label{eq2.6}
w=\log\left(\frac{z}{\ell-z}\right),
\end{equation}
where $\ell$ is the length of the region $A$, and we have assumed the origin to coincide with the left boundary between $A$ and $B$. Under this transformation, see Figure \ref{fig:ii}, the partition function is given by
\begin{equation} \label{eq2.7}
Z_n^{\alpha\beta}=\mathrm{Tr}_{\mathcal{H}_{\alpha\beta}}\left[q^{n\left(L_0-\frac{c}{24}\right)}\right],
\end{equation}
where $q=e^{i2\pi\tau}$, with the modular parameter $\tau=i\frac{\pi}{W}$, and the width of the annulus $W=2\log\frac{\ell}{\epsilon}+O(\epsilon)$. The operators $L_n$ are the Virasoro generators of the CFT. The subscript $\mathcal{H}_{\alpha\beta}$ in eq.\eqref{eq2.7} is to denote that the trace is over the Hilbert space $\mathcal{H}_{\alpha\beta}$. The Hilbert space $\mathcal{H}_{\alpha\beta}$ is usually a priori unknown, however, we may exploit the open/closed string duality to evaluate eq.\eqref{eq2.7}. Using this duality, that is the S-transform of eq.\eqref{eq2.7}, the partition function $Z_n$ equals the matrix element
\begin{equation}\label{eq2.8}
Z_n^{\alpha\beta}=\langle\left.\alpha\right|\tilde{q}^{\frac{1}{2n}\left(L_0+\bar{L}_0-\frac{c}{12}\right)}\left|\beta\right.\rangle,
\end{equation}
on the cylinder, where $\tilde{q}=e^{-2W}$. In the limit $\epsilon\to 0$, we have $\tilde{q}\to 0$, for future use, we will call this limit the asymptotic limit. The state $\left|\alpha\right.\rangle$, and $\left|\beta\right.\rangle$ are the Cardy states \cite{cardy1989boundary} respecting the conformal boundary conditions $\alpha$, and $\beta$ respectively. The R\'enyi entropies are then given by
\begin{equation}\label{eq2.9}
S_n=\frac{1}{1-n}\log\left(\frac{Z^{\alpha\beta}_n}{\left(Z^{\alpha,\beta}_1\right)^n}\right).
\end{equation} 
Let's now further assume that the boundary conditions $\alpha$, and $\beta$ both preserve the symmetry under scrutiny in addition to the conformal symmetry. It has been noted in \cite{di2023boundary}, that for such boundary conditions, the partition function $Z_n$ is already diagonalised into the local charge sectors $j$, this may be understood as follows. Let us begin by considering that both the boundary conditions preserve the same extended symmetry $W$. The Hilbert space $\mathcal{H}_{\alpha\beta}$ of the boundary theory in this case decomposes into the charge sectors $\mathcal{H}_j$ of the zero mode of the current algebra $W(z)$. The partition function in the open string channel may be written as
\begin{equation} \label{eq2.10}
Z_n^{\alpha\beta}=\sum_{j}n_j\chi_{j}(q),
\end{equation}
where $\chi_j(q)$ is just the trace over the Hilbert space sector $\mathcal{H}_j$ and $n_j$ is the non-trivial multiplicity of $\mathcal{H}_j$ in $\mathcal{H}_{\alpha\beta}$. This implies that the partition function $Z^{\alpha\beta}_n$ is already diagonalised in the local charge sectors $j$. Now if the conformal boundary conditions $\alpha$ and $\beta$ only preserve a subgroup of the symmetries preserved by either of the boundary conditions, the Hilbert space $\mathcal{H}_{\alpha\beta}$ still decomposes into the charge sectors corresponding to the preserved symmetries.
\section{Massless Dirac Fermion} \label{section3}
In this section, we study the symmetry resolution of massless Dirac fermions and $Z_2$ gauged massless Dirac fermions using the BCFT approach. This approach allows us to study different spin sectors and their contribution to the symmetry-resolved entanglement.
\subsection{Dirac Fermion} \label{section3.1}
The massless Dirac fermion has central charge $c=1$, and it is given by the euclidean action
\begin{equation} \label{eq3.1}
S=\frac{1}{4\pi}\int\mathrm{d}^2z\bar{\Psi}\slashed{\partial}\Psi,
\end{equation}
where $\gamma^0=\sigma^1$, and $\gamma^1=\sigma^2$, $\sigma^i$ being the Pauli matrices. This theory possesses a $U(1)$ symmetry under the transformation $\Psi \to e^{i\theta}\Psi$, with the corresponding conserved current $J^{\mu}_{V}=\bar{\Psi}\gamma^{\mu}\Psi$, known as the Vector current. There exists another $U(1)$ symmetry in the theory with the corresponding conserved current $J^{\mu}_{A}=\bar{\Psi}\gamma^{\mu}\sigma^{3}\Psi$, this is the Axial current. We will study the symmetry decomposition of entanglement with respect to both the $U(1)$ symmetries.  The action $S$, may be written in terms of two Weyl fermions as
\begin{equation} \label{eq3.2}
S=\frac{1}{4\pi}\int\mathrm{d}^2 z\psi^{\dagger}\partial_{\bar{z}}\psi+\bar{\psi}^{\dagger}\partial_{\bar{z}}\bar{\psi},
\end{equation}
where $\psi$, and $\bar{\psi}$ are holomorphic and anti-holomorphic fermions respectively. The holomorphic current $J(z)$ of the theory, corresponding to the symmetry transformation $\psi \to e^{i\theta}\psi$, is given by $J(z)=:\psi^{\dagger}\psi:(z)$. We take the following mode expansions in the $\mathrm{NS}$ sector (i.e anti-periodic boundary conditions on the cylinder)
\begin{align}
\psi(z)&=\sum_{n\in\mathbb{Z}+1/2} z^{-n-1/2}\psi_{n}, \label{eq3.3}\\
T(z)&=\sum_{n\in \mathbb{Z}} z^{-n-2}L_{n}, \label{eq3.4}\\
J(z)&=\sum_{n\in \mathbb{Z}} z^{-n-1}J_{n}, \label{eq3.5}
\end{align}
where, $T(z)$ is the holomorphic stress energy tensor. We have the following operator algebra for the current mode operators
\begin{align}
[J_m,J_n]&=m\delta_{m+n,0}, \qquad [J_m,L_n]=mJ_{m+n},\label{eq3.6}\\
[J_m,\psi_n]&=\psi_{n+m},  \qquad [J_m,\psi^{\dagger}_n]=-\psi^{\dagger}_{n+m}. \label{eq3.7}
\end{align}

As discussed in Section \ref{section2}, to study symmetry-resolved entanglement we need to evaluate the BCFT partition function with symmetry preserving boundary. To evaluate the partition function in the closed string channel we need the corresponding Cardy boundary states in eq.\eqref{eq2.8} first. The Vector and Axial $U(1)$ symmetries have an anomaly between them and hence a boundary condition cannot be constructed which preserves both symmetries. The symmetry preserving boundary conditions for Vector and Axial $U(1)$ symmetries on the real line ($z=\bar{z}$) are given by
\begin{align}
\psi(z)=e^{i\theta}\bar{\psi}(\bar{z}) \qquad V, \label{eq3.8}\\
\psi(z)=e^{i\theta}\bar{\psi}^{\dagger}(\bar{z}) \qquad A, \label{eq3.9}
\end{align}
where the phase $\theta \in \left[0,2\pi\right)$. The notation $V$, and $A$ above stand for Vector, and Axial respectively. The boundary states in the $\mathrm{NS}$ sector are known, see for example \cite{smith2020boundary}. We have however presented a brief calculation of the boundary states in Appendix \ref{B}. A few good reviews on the construction of boundary states are \cite{recknagel2013boundary, gaberdiel2001conformal,blumenhagen2009introduction}. These boundary states are given by
\begin{equation}
\left| V/A,\theta\right\rangle_{\mathrm{NS}}=\sum_{p\in\mathbb{Z}}e^{-i\pi p^2+ip,\theta}\left|\left| p,\theta\right\rangle\right\rangle_{V/A,\mathrm{NS}}, \label{eq3.10}
\end{equation} 
where $\theta$ is the same as in the boundary conditions eq.\eqref{eq3.8}-\eqref{eq3.9}, $p$ corresponds to $J_0$ eigenvalues, and $\left|\left| p,\theta\right\rangle\right\rangle_{V/A,\mathrm{NS}}$ are the Ishibashi boundary states \cite{ishibashi1989boundary}. The Ishibashi boundary states are given by
\begin{align}
\left|\left| p,\theta\right\rangle\right\rangle_{V,\mathrm{NS}}&=e^{\sum_{n=1}^{\infty}\frac{1}{n}J_{-n}\bar{J}_{-n}}\left|p,\bar{p}_V\right\rangle,\label{eq3.11}\\
\left|\left| p,\theta\right\rangle\right\rangle_{A,\mathrm{NS}}&=e^{-\sum_{n=1}^{\infty}\frac{1}{n}J_{-n}\bar{J}_{-n}}\left|p,\bar{p}_A\right\rangle \label{eq3.12}
\end{align} 
where the antiholomorphic charge $\bar{p}_V=p$, and $\bar{p}_A=-p$ for the Vector and Axial boundary conditions respectively. The states $\left|p,\bar{p}\right\rangle$ in \eqref{eq3.11}-\eqref{eq3.12} satisfy
\begin{align}
&J_n\left|p,\bar{p}\right\rangle=0, \\
&\bar{J}_n\left|p,\bar{p}\right\rangle=0, \qquad \forall n \geq 1,
\end{align}
for either boundary state.

Let's first give the boundary partition function for the same type of boundary condition imposed on both sides in eq.\eqref{eq2.8}. The boundary partition function $Z_{n}^{\Delta\theta,AA/VV}=\left\langle V/A,\theta\right|\tilde{q}^{\frac{1}{2n}\left(L_0+\bar{L}_0-\frac{c}{12}\right)}\left| V/A,\theta'\right\rangle$, where $\Delta\theta=\theta-\theta'$, is the same in both cases, so we drop the superscripts in our notation for what follows. The partition function is then given by
\begin{align}
Z_{n}^{\left(\Delta\theta\right)}=\sum_{p\in\mathbb{Z}}e^{ip\Delta\theta}\frac{\tilde{q}^{p^2/2n}}{\eta\left(-\frac{1}{n\tau}\right)}=\frac{\vartheta_{4}\left(\frac{\Delta\theta}{2\pi},-\frac{1}{n\tau}\right)}{\eta\left(-\frac{1}{n\tau}\right)}, \label{eq3.13}
\end{align}
where $\eta$ is the Dedekind eta function, $\vartheta_i$ are the Jacobi theta functions and the moduli $\tau$ is given below eq.\eqref{eq2.7}. To write this partition function in the desired form, the open string channel, we may use the Poisson resummation formula to rewrite it as
\begin{equation} \label{eq3.14}
Z_{n}^{\left(\Delta\theta\right)}=\sum_{p\in \mathbb{Z}}\frac{q^{\frac{n}{2}\left(\frac{\Delta\theta}{2\pi}+p\right)^2}}{\eta(n\tau)}.
\end{equation}
The partition function above is diagonal in the local charge sectors $Q$, where $Q\in\{\frac{\Delta\theta}{2\pi}+n|n\in\mathbb{Z}\}$.

The partition function of the anti-periodic Dirac fermion on the cylinder (i.e. the Ramond sector), is straightforward to obtain in the closed channel by insertion of the operator $(-1)^{F+\bar{F}}$, where $F(\bar{F})$ counts the number of holomorphic(anti-holomorphic) fermions, inside the matrix element in eq.\eqref{eq2.8}. Since all the states in eq.\eqref{eq3.11}-\eqref{eq3.12} are even under $(-1)^{F+\bar{F}}$, we have the same partition function for the Ramond sector as well. This implies the same entanglement spectrum in both sectors. In \cite{foligno2023entanglement}, it was shown that the symmetry resolved entanglement of Dirac fermion on a torus for a single interval differed in different spin sectors, however, the spin sector dependent part goes to zero in the zero temperature limit, consistent with our results. The entanglement entropies, using eq.\eqref{eq3.13}, are given by
\begin{align}
S_n=&\frac{1+n}{n}\frac{W}{12}+\frac{1}{1-n}\sum_{k=1}^{\infty} \log\left(\frac{\left(1-\tilde{q}^k\right)^n}{1-\tilde{q}^{k/n}}\right) +\frac{1}{1-n}\log\left(\frac{\sum_{p\in\mathbb{Z}}e^{ip\Delta\theta}\tilde{q}^{p^2/2n}}{\left(\sum_{p\in\mathbb{Z}}e^{ip\Delta\theta}\tilde{q}^{p^2/2}\right)^n}\right), \label{eq3.15}\\
\begin{split}
S_1=&\frac{W}{6}-\sum_{k=1}^{\infty}\left(\log\left(1-\tilde{q}^k\right)-\frac{k\log(\tilde{q})\tilde{q}^{k}}{1-\tilde{q}^{k}}\right)\\
&\qquad +\left[\frac{1}{2}\log(\tilde{q})\frac{\sum_{p\in\mathbb{Z}}p^2e^{ip\Delta\theta}\tilde{q}^{p^2/2}}{\sum_{p\in\mathbb{Z}}e^{ip\Delta\theta}\tilde{q}^{p^2/2}}+\log\left(\sum_{p\in\mathbb{Z}}e^{ip\Delta\theta}\tilde{q}^{p^2/2}\right)\right]. \label{eq3.16}
\end{split}
\end{align}
We have well known $\frac{c}{3}\log(\ell/\epsilon)$ leading order term in eq.\eqref{eq3.16}. The subleading order contribution is of the order $\tilde{q}^{1/n}$ in eq.\eqref{eq3.15}-\eqref{eq3.16}, which vanishes in the Asymptotic limit $\epsilon\to 0$. We also note that there is no Affleck-Ludwig boundary entropy term \cite{affleck1991universal} in this case. Since for many cases, including the single interval case, the entanglement entropies of the Dirac fermion are known to match with that of the self-dual compact boson \cite{headrick2013bose}, we note that however, this is not the case for boundary entropy term as self-dual compact boson has non-vanishing boundary entropy \cite{di2023boundary}. We mention that the two theories are not dual to each other, as discussed further in the next subsection. Finally, the symmetry resolved entanglement entropies are given by
\begin{align}
\begin{split}
S_n(Q)=&\frac{1}{1-n}\left[\log\left(\frac{q^{nQ^2/2}}{\eta(n\tau)}\right)-n\log\left(\frac{q^{Q^2/2}}{\eta(\tau)}\right)\right]\\
=& \frac{1+n}{n}\frac{W}{12}-\frac{1}{2}\log\left(\frac{W}{\pi}\right)+\frac{1}{2(1-n)}\log(n)+\frac{1}{1-n}\sum_{k=1}^{\infty}\log\left[\frac{\left(1-\tilde{q^k}\right)^n}{1-\tilde{q}^{k/n}}\right], \label{eq3.17}
\end{split}\\
S_{1}(Q)=&\frac{W}{6}-\frac{1}{2}\log\left(\frac{W}{\pi}\right)-\frac{1}{2}-\sum_{k=1}^{\infty}\left[\log\left(1-\tilde{q}^k\right)-k\log\left(\tilde{q}\right)\frac{\tilde{q}^{k/n}}{1-\tilde{q}^{k/n}}\right]. \label{eq3.18}
\end{align}
The leading order term is again $\frac{c}{3}\log\left(\ell/\epsilon\right)$, and the sub-leading term in eq.\eqref{eq3.18} is the familiar $\log\left(\log(\ell/\epsilon)\right)$ term for $U(1)$ symmetry resolved entanglement. This is exactly the results obtained in \cite{di2023boundary} for free compact boson. We note that while the entanglement entropy of Dirac fermion and self-dual free compact boson differ at $O(1)$, the symmetry resolved entanglement depends upon the representation of the $U(1)$ symmetry group and is indeed the same in both the theories. Finally, the fluctuation entropy in eq.\eqref{eq2.5} is readily obtained and to the leading order has a $\log\left(\log(\ell/\epsilon)\right)$ dependence.

Finally imposing different classes of boundary conditions on each boundary leads to the partition function $Z_1$ (at $\theta=0$)
\begin{equation} \label{eq3.19}
\begin{split}
Z_{1}^{AV}=&\tilde{q}^{-1/24}\prod_{k=0}^{\infty}\left(1+\tilde{q}^k\right)^{-1}=\frac{\eta\left(-\frac{1}{\tau}\right)}{\eta\left(-\frac{2}{\tau}\right)}\\
=&\sqrt{2}\frac{\eta\left(\tau\right)}{\eta\left(\frac{\tau}{2}\right)}.
\end{split}
\end{equation}
The resulting BCFT is anomalous due to the $\sqrt{2}$ factor in the partition function above which reflects the presence of a single unpaired Majorana fermion. Consequently, these boundary conditions are not mutually compatible. 
\subsection{Modular invariant $Z_2$-gauged Dirac fermion} \label{section3.2}
In this subsection, we consider the modular invariant massless Dirac fermions. These fermions are gauged by the operator $(-1)^{F+\bar{F}}$. In this case, we also need to consider the Ramond sector contribution as well. The calculations in the Neveu-Schwarz sector follow from the last subsection. The Ramond sector calculations are a little more involved due to the presence of the fermion zero modes. We have the following mode expansion of the fermion field $\psi(z)$ in the Ramond sector
\begin{equation} \label{eq3.20}
\psi(z)=\sum_{n\in\mathbb{Z}}z^{-n-1/2}\psi_n.
\end{equation}
The current mode operators in eq.\eqref{eq3.5} may be expressed in terms of the fermion mode operators of eq.\eqref{eq3.20} as
\begin{equation} \label{eq3.21}
J_n=-\sum_{k\geq 0}\psi^{\dagger}_{n-k}\psi_{k}+\sum_{k\leq -1}\psi_{k}\psi^{\dagger}_{n-k}+\frac{1}{2}\delta_{n,0}.
\end{equation}
We also have the following algebra for the zero modes
\begin{equation} \label{eq3.22}
\left\{\psi_0^{\dagger},\psi_0\right\}=1,\qquad \left[J_0,\psi_0\right]=\psi_0, \qquad \left[J_0,\psi^{\dagger}_0\right]=-\psi^{\dagger}_0.
\end{equation}
Using the zero mode algebra we define the degenerate ground states $\left|\sigma\right\rangle$, and $\left|\rho\right\rangle$, satisfying the following equations
\begin{equation} \label{eq3.23}
\begin{split}
&\psi_0\left|\sigma\right\rangle=0,\qquad \psi^{\dagger}_0\left|\rho\right\rangle=0,\\
&\psi_0\left|\sigma\right\rangle=\left|\rho\right\rangle,\qquad \psi^{\dagger}_0\left|\rho\right\rangle=\left|\sigma\right\rangle.
\end{split}
\end{equation}
The ground states carry non-zero $U(1)$ charge and it follows from definitions in eq.\eqref{eq3.23} and eq.\eqref{eq3.22} that they satisfy
\begin{equation} \label{eq3.24}
J_0\left|\sigma\right\rangle=\frac{1}{2}\left|\sigma\right\rangle, \qquad J_0\left|\rho\right\rangle=-\frac{1}{2}\left|\rho\right\rangle.
\end{equation}
The stress-energy tensor may be expressed in a similar fashion. We first write the Weyl fermion in terms of two holomorphic Majorana fermions by
\begin{equation} \label{eq3.25}
\psi(z)=\frac{1}{\sqrt{2}}\left(\psi^{1}(z)+i\psi^{2}(z)\right).
\end{equation} 
The stress-energy tensor $T(z)$ is then just the sum of the two stress-energy tensors $T^{i}(z)$, where the $L^{i}_n$ modes are given in terms of the fermion modes in eq.\eqref{eq3.20} as
\begin{equation} \label{eq3.26}
L^{i}_n=\frac{1}{2}\sum_{s\geq-1}\left(s+\frac{1}{2}\right)\psi^{i}_{n-s}\psi^{i}_{s}-\frac{1}{2}\sum_{s\leq-2}\left(s+\frac{1}{2}\right)\psi^{i}_{s}\psi^{i}_{n-s}+\frac{3}{16}\delta_{n,0}.
\end{equation}
The Virasoro operator $L_0$ acts on the ground states $\left|\sigma\right\rangle$, and $\left|\rho\right\rangle$ to give
\begin{equation} \label{eq3.27}
L_0\left|\sigma\right\rangle=\frac{1}{8}\left|\sigma\right\rangle, \qquad L_0\left|\rho\right\rangle=\frac{1}{8}\left|\rho\right\rangle
\end{equation}
The construction of the boundary states in the Ramond sector has been discussed in Appendix \ref{B}. These boundary states are given by
\begin{align}
\left|V,\theta\right\rangle_{\mathrm{R}}=&\tilde{a}_0\sum_{p\in\mathbb{Z}}e^{i\theta p+i\pi p^2/2}\left|\left|\lambda_p,\lambda_p\right\rangle\right\rangle_{V,\mathrm{R}},\label{eq3.28}\\
\left|A,\theta\right\rangle_{\mathrm{R}}=&\tilde{a}_0\sum_{p\in\mathbb{Z}}e^{i\theta p+i\pi p^2/2}\left|\left|\lambda_p,-\lambda_p\right\rangle\right\rangle_{A,\mathrm{R}},\label{eq3.29}
\end{align}
where the $U(1)$ charge $\lambda_p=p+\frac{1}{2}$ and the Ishibashi states are given as in eq.\eqref{eq3.11}-\eqref{eq3.12} using eq.\eqref{eq3.21}. The base ground states in Vector and Axial boundary CFT are chosen to be $\left|\sigma,\bar{\sigma}\right\rangle$ and $\left|\sigma,\bar{\rho}\right\rangle$ respectively, see Appendix \ref{B}.

The boundary states for the modular invariant Dirac fermion is given by
\begin{align}
\left|V,\theta\right\rangle=&{g_V}\sum_{p\in\mathbb{Z}}e^{i\theta p}\left|\left|p,p\right\rangle\right\rangle_{V,\mathrm{NS}}+g_V\sum_{p\in\mathbb{Z}+\frac{1}{2}}e^{i\theta p+i\pi \frac{(p-1/2)^2}{2}}\left|\left|p,p\right\rangle\right\rangle_{V,\mathrm{R}}, \label{eq3.30}\\
\left|A,\theta\right\rangle=&g_{A}\sum_{p\in\mathbb{Z}}e^{i\theta p}\left|\left|p,-p\right\rangle\right\rangle_{A,\mathrm{NS}}+g_{A}\sum_{p\in\mathbb{Z}+\frac{1}{2}}e^{i\theta p+i\pi \frac{(p-1/2)^2}{2}}\left|\left|p,-p\right\rangle\right\rangle_{A,\mathrm{R}}, \label{eq3.31}
\end{align}
where we have imposed $e^{-i\theta/2}\tilde{a}_0=a_0=g_{A/V}$ in order to satisfy the Cardy consistency conditions \cite{cardy1989boundary}. In order to compute the partition function we must find and project out the $(-1)^{F+\bar{F}}$ odd states. The closed channel string partition function to be evaluated is
\begin{equation} \label{eq3.32}
Z_{n}^{\Delta\theta}=\frac{1}{2}\left\langle B_1,\theta'\right| \left(1+(-1)^{F+\bar{F}}\right)\tilde{q}^{\frac{1}{n}\left(L_0-\frac{1}{24}\right)}\left| B_2,\theta\right\rangle,
\end{equation} 
where $B_1,B_2\in \{V,A\}$. To evaluate $(-1)^{F+\bar{F}}$ projection, we must have a representation of $(-1)^{F+\bar{F}}$ in the Ramond sector. Since we have already constructed the action of $J_0$, and $\bar{J}_0$ on the ground states, we may use the $(-1)^{J_0+\bar{J}_0}$ representation. The states for both boundary conditions in the Neveu-Schwarz sectors are $(-1)^{F+\bar{F}}$ even states. In the Ramond sector, all the states are $(-1)^{F+\bar{F}}$ even and odd states for the Vector and Axial boundary conditions respectively. We therefore have the following BCFT partition function for the same class of boundary conditions imposed on both boundaries
\begin{align}
\begin{split} 
Z^{\left(\Delta\theta\right)}_{n,V}=&g_{V}^2\tilde{q}^{-\frac{1}{24n}}\sum_{p\in\mathbb{Z}}e^{ip\Delta\theta}\frac{\tilde{q}^{\frac{p^2}{2n}}}{\eta\left(-\frac{1}{n\tau}\right)}+g_{V}^2\tilde{q}^{-\frac{1}{24n}}\sum_{p\in\mathbb{Z}+\frac{1}{2}}e^{ip\Delta\theta}\frac{\tilde{q}^{\frac{1}{2n}\left(p^2-\frac{1}{4}\right)+\frac{1}{8n}}}{\eta\left(-\frac{1}{n\tau}\right)}\\
=&g_{V}^2\tilde{q}^{-\frac{1}{24n}}\sum_{p\in\mathbb{Z}}e^{i\frac{p\Delta\theta}{2}}\frac{\tilde{q}^{\frac{p^2}{8n}}}{\eta\left(-\frac{1}{n\tau}\right)} \label{eq3.33}
\end{split}\\
Z^{\left(\Delta\theta\right)}_{n,A}=&g_{A}^2\tilde{q}^{-\frac{1}{24n}}\sum_{p\in\mathbb{Z}}e^{ip\Delta\theta}\frac{\tilde{q}^{\frac{p^2}{2n}}}{\eta\left(-\frac{1}{n\tau}\right)}, \label{eq3.34}
\end{align}
where $g_{V/A}$ is the Affleck-Ludwig g-factor \cite{affleck1991universal} also appearing in eq.\eqref{eq3.30}-\eqref{eq3.31}, for the Vector boundary condition $g_{V}$ equals $\sqrt{\frac{1}{2}}$ and for Axial boundary conditions $g_{A}=1$.

There exists a duality between the modular invariant Dirac fermion and free compact boson at compactfication radius $R=1$, following the conventions of \cite{francesco2012conformal}. The bosonisation relations are 
\begin{equation} \label{eq3.35}
\begin{split}
\psi(z)&\sim e^{i\varphi(z)}\\
\bar{\psi}(\bar{z})&\sim e^{i\bar{\varphi}(\bar{z})}.
\end{split}
\end{equation}
We would expect the two theories to have the same entanglement spectrum \cite{headrick2013bose}. The issue of entanglement spectrum between the two theories on the torus has been further discussed in \cite{lokhande2015modular, mukhi2018entanglement}. We note from  eq.\eqref{eq3.33}-\eqref{eq3.34}, that the partition function of the two theories matches exactly, see ref.\cite{di2023boundary} for compact boson results. The Vector boundary condition partition function for the modular invariant Dirac fermion matches with the Neumann boundary condition partition function for the free compact boson and the Axial boundary condition partition function matches with the Dirichlet boundary condition partition function. The boundary operator content of the two theories matches exactly, as one would expect from eq.\eqref{eq3.8}-\eqref{eq3.9}, and  eq.\eqref{eq3.35}. We also note that the Axial boundary condition result matches with the corresponding results of the last subsection. The formal reason for this result is due to the fact that all the states in the Ramond sector are $(-1)^{F+\bar{F}}$ odd, and hence don't contribute.

The partition function in the open string channel for the Vector boundary condition is given by
\begin{equation} \label{eq3.36}
Z^{\left(\Delta\theta\right)}_{n,V}=\sum_{p\in\mathbb{Z}}\frac{q^{\frac{n}{2}\left(2p+\frac{\Delta\theta}{2\pi}\right)^2}}{\eta\left(n\tau\right)}.
\end{equation}
The corresponding partition function for the Axial boundary conditions is given by eq.\eqref{eq3.13}. The entanglement entropies for the Vector boundary condition are given by
\begin{align}
S_n=&\frac{1+n}{n}\frac{W}{12}+\log\left(\frac{1}{2}\right)+\sum_{k=1}^{\infty} \log\left(\frac{\left(1-\tilde{q}^k\right)^n}{1-\tilde{q}^{k/n}}\right) +\frac{1}{1-n}\log\left(\frac{\sum_{p\in\mathbb{Z}}e^{i\frac{p}{2}\Delta\theta}\tilde{q}^{p^2/8n}}{\left(\sum_{p\in\mathbb{Z}}e^{i\frac{p}{2}\Delta\theta}\tilde{q}^{p^2/8}\right)^n}\right), \label{eq3.37}\\
\begin{split}
S_1=&\frac{W}{6}+\log\left(\frac{1}{2}\right)-\sum_{k=1}^{\infty}\left(\log\left(1-\tilde{q}^k\right)-\frac{k\log(\tilde{q})\tilde{q}^{k}}{1-\tilde{q}^{k}}\right)\\
&\qquad +\left[\frac{1}{8}\log(\tilde{q})\frac{\sum_{p\in\mathbb{Z}}p^2e^{i\frac{p}{2}\Delta\theta}\tilde{q}^{p^2/8}}{\sum_{p\in\mathbb{Z}}e^{i\frac{p}{2}\Delta\theta}\tilde{q}^{p^2/8}}+\log\left(\sum_{p\in\mathbb{Z}}e^{i\frac{p}{2}\Delta\theta}\tilde{q}^{p^2/8}\right)\right]. \label{eq3.38}
\end{split}
\end{align}
The entanglement entropies above show a behaviour similar to eq.\eqref{eq3.15}-\eqref{eq3.16}, except for the non-vanishing Affleck-Ludwig boundary entropy term. The entanglement entropy of the Axial boundary condition is the same as in eq.\eqref{eq3.15}-\eqref{eq3.16}. Finally, the symmetry-resolved entanglement entropies for both boundary conditions are given by eq.eq.\eqref{eq3.17}-\eqref{eq3.18}, the only difference being the charge spectrum. The charge spectrum for $\mathrm{V}$ is $Q\in\{2p+\frac{\Delta\theta}{2\pi}|p\in\mathbb{Z}\}$, while for $\mathrm{A}$ the charge spectrum is $Q\in\{p+\frac{\Delta\theta}{2\pi}|p\in\mathbb{Z}\}$.

Finally imposing different boundary conditions on either end leads to the partition function
\begin{equation} \label{eq3.39}
Z_n^{\mathrm{AV}}=Z_n^{\mathrm{VA}}=\frac{\eta(n\tau)}{\eta\left(\frac{n\tau}{2}\right)},
\end{equation}
we see that there is no anomaly for $\mathrm{AV}$ or $\mathrm{VA}$ boundary conditions for the modular invariant Dirac fermion as opposed to the results of the last subsection. In the open channel, the mixed boundary conditions imply that the current $J(z)$ admits mode expansion in terms of half integer modes $J_n$, where $n\in \{\mathbb{Z}+1/2\}$ \cite{blumenhagen2009introduction}. Consequently the mixed boundary condition partition function is just a single twisted $U(1)$ character built on a primary with conformal dimension $h=\frac{1}{16}$. There is a $Z_2$ symmetry corresponding to the Hilbert space sectors containing states constructed by even and odd number of $J_{-n}$ modes, where $n>0$, respectively. The symmetry resolution of entanglement entropy with respect to this symmetry has been studied in \cite{di2023boundary} for free compact boson, and applies similarly in the present study as well. 
\section{N Dirac fermions} \label{section4}
In this section, we study the entanglement spectrum and its symmetry resolution for $N$ massless Dirac fermion. We will consider the boundary conditions preserving $U(1)^N$ symmetries, and study entanglement for $U(1)^M$ subgroups, where $M\leq N$.

This theory possesses a $SO(2N)\times SO(2N)$ symmetry. The boundary theory of $N$ massless Dirac fermions preserving all chiral $U(1)^N$ symmetries have been studied in \cite{smith2020boundary, boyle2021boundary}. We first give a brief introduction of these results in the subsection \ref{section4.1}. We then proceed to calculate the entanglement spectrum and its symmetry resolution for $U(1)^M$ subgroups in the subsection \ref{section4.2}.
\subsection{Boundary states for $U(1)^N$ chiral symmetries} \label{section4.1}
In this subsection, we briefly review the results of Ref.\cite{smith2020boundary}. To begin our discussion, let us assign fermions with holomorphic charges $Q_i$ and anti-holomorphic charges $\bar{Q}_i$, where $i\in\{1,2,\cdots,N\}$ and denotes the $i^{th}$ fermion. We may construct with these fixed charges the currents $\mathcal{J}=Q_iJ_i$ and $\bar{\mathcal{J}}=\bar{Q}_i\bar{J}_i$. The currents $\mathcal{J}$, and $\bar{\mathcal{J}}$ correspond to a $U(1)$ symmetry. Let's now further assume that we have $N$ such symmetries. Let us denote the charges corresponding to these symmetries by $Q_{\alpha,i}$ where $\alpha$ now stands for a $U(1)$ symmetry among the $N$ $U(1)$ symmetries. If these charges satisfies the conditions
\begin{equation} \label{eq4.1}
\sum_{i}Q_{\alpha,i}Q_{\beta,i}=\sum_{i}\bar{Q}_{\alpha,i}\bar{Q}_{\alpha,i},
\end{equation}
then the corresponding chiral $U(1)^N$ is free from mixed 't hooft anomalies. The $U(1)^N$ symmetry preserves the currents,
\begin{equation} \label{eq4.2}
\mathcal{J}_{\alpha}=Q_{\alpha,i}J_i, \qquad \bar{\mathcal{J}}_{\alpha}=\bar{Q}_{\alpha,i}\bar{J}_i.
\end{equation}
Each such distinct $U(1)^N$ subgroup of $SO(2N)\times SO(2N)$ symmetry of the theory is distinguished by the orthogonal matrix
\begin{equation} \label{eq4.3}
R_{ij}=\bar{Q}^{-1}_{i,\alpha}Q_{\alpha,j}.
\end{equation}
An $N$-dimensional lattice, denoted $\Lambda[R]$, is also associated with the $U(1)^N$ symmetry. The lattice $\Lambda[R]\in\mathbb{Z}^N$, where the space $\mathbb{Z}^N$ represents the space of holomorphic $U(1)$ charges carried by $N$ fermions we started with. The lattice $\Lambda[R]$ is the collection of all the allowed holomorphic charges carried by $U(1)^N$ chiral currents. Formally, the lattice $\Lambda[R]$ is given by
\begin{equation} \label{eq4.4}
\Lambda[R]=\{\lambda\in \mathbb{Z}^N|\bar{\lambda}=-R\lambda\in \mathbb{Z}^N\},
\end{equation}
where $\bar{\lambda}$ is the corresponding anti-holomorphic part of the chiral charge. The boundary states corresponding to the boundary conditions preserving the given $U(1)^N$ symmetry must satisfy
\begin{equation} \label{eq4.5}
\left(R_{i,j}J_j+\bar{J}_i\right)\left|B\right\rangle=0.
\end{equation}
The consistent set boundary states satisfying these conditions are given by
\begin{equation} \label{eq4.6}
\left|a,R\right\rangle=\sum_{\lambda\in \Lambda[R]}a_{\lambda}\left|\left|\lambda,-R\lambda,R\right\rangle\right\rangle,
\end{equation}
where the Ishibashi states $\left|\left|\lambda,-R\lambda,R\right\rangle\right\rangle$ are defined as
\begin{equation} \label{eq4.7}
\left|\left|\lambda,-R\lambda,R\right\rangle\right\rangle=e^{-\sum_{n=1}^{\infty}\frac{1}{n}R_{ij}J_{i,-n}\bar{J}_{j,-n}}\left|\lambda,-R\lambda\right\rangle.
\end{equation}
The complex coefficients in eq.\eqref{eq4.6} are $a_{\lambda}=a_{0}e^{i\theta\cdot\lambda+i\gamma_R\left(\lambda\right)}$, where $\theta_i$ are the phases similar to the ones appearing in eq.\eqref{eq3.8}-\eqref{eq3.9}. The factor $e^{i\gamma_R\left(\lambda\right)}$ appears in solutions to the Cardy-Lewellen sewing conditions. Since its exact form is not required here, we don't discuss it further. The partition function for the boundary conditions preserving the same $U(1)^N$ chiral symmetry is given by
\begin{equation} \label{eq4.8}
\begin{split}
Z_{R}^{\Delta\theta}=&g_{R}^{2}\sum_{\lambda\in\Lambda[R]}\frac{\tilde{q}^{\frac{\lambda^2}{2}}e^{\Delta\theta\cdot\lambda}}{\eta\left(-\frac{1}{\tau}\right)^N}\\
=&\frac{1}{\eta\left(\tau\right)^N}\sum_{\mu\in\Lambda^{*}[R]}q^{\frac{1}{2}\left(\mu+\Delta\theta\right)^2},
\end{split}
\end{equation}
where the Affleck-Ludwig g-factor is $g_R=a_0=\sqrt{\mathrm{Vol}\left(\Lambda[R]\right)}$, the volume of the primitive cell in $\Lambda[R]$, was used to satisfy the cardy conditions. The first equality in eq.\eqref{eq4.8} gives the partition function in the closed string channel, while the second equality gives the partition function in the open string channel. The lattice $\Lambda^{*}[R]$ is dual of the lattice $\Lambda[R]$. The partition, as written in the open string channel, is diagonal in the $U(1)^N$ charge sectors and the charges are given by $\mu+\frac{\Delta\theta}{2}$.

Let us now proceed to discuss the case where boundary conditions preserving different $U(1)^N$ symmetries are imposed on the two boundaries such that a $U(1)^M$ subgroup of common to both $U(1)^N$ symmetries is still preserved. The generic partition function, in closed string channel, in this case, given by
\begin{equation} \label{eq4.9}
Z^{\Delta\theta}_{R',R}=g_{R'}g_{R}\tilde{q}^{-\frac{N}{24}}\prod_{m=1}^{\infty}\frac{1}{\det\left(\mathbb{I}-\tilde{q}^N R'^{T}\cdot R\right)}\sum_{\lambda\in\tilde{\Lambda}[R',R]}e^{i\left(\Delta\theta+\pi s\right)\cdot \lambda}\tilde{q}^{\frac{\lambda^2}{2}},
\end{equation}
where the matrices $R'$, and $R$ correspond to either boundary condition. The lattice $\tilde{\Lambda}[R',R]$ is a sub-lattice common to both the lattices $\Lambda[R']$, and $\Lambda[R]$. Mathematically it is given by $\left\{\lambda\in\mathbb{Z}^n|R\lambda=R'\lambda\in\mathbb{Z}^n\right\}$. The factor $e^{is}$ in eq.\eqref{eq4.9} results from $e^{i\gamma_{R}(\lambda)-i\gamma_{R'}(\lambda)}=e^{is}$, for some $s\in\tilde{\Lambda}[R',R]$. The orthogonal matrix $R'^{T}\cdot R$ must have $M$ eigenvalues with value $+1$ corresponding to the preserved $U(1)^M$ symmetry. The remaining $N-M$ eigenvalues are either $-1$ or appear in the conjugate pair $e^{\pm it_j}$, where $t_j\in (0,\pi)$. Each $-1$ eigenvalue reflects a remnant $Z_2$ symmetry in the theory. Let us denote the number of conjugate pair eigenvalues $e^{\pm it_j}$ by $p$. The partition function in eq.\eqref{eq4.9} may now be rewritten more conveniently as
\begin{equation} \label{eq4.10}
\begin{split}
Z^{\Delta\theta}_{R',R}=g_{R'}g_{R}\left(\frac{1}{\eta\left(-\frac{1}{\tau}\right)^M}\right)\left(\frac{\eta\left(-\frac{2}{\tau}\right)}{\eta\left(-\frac{1}{\tau}\right)}\right)^{N-M-2p}&\left(\frac{1}{\eta\left(-\frac{1}{\tau}\right)^{2p}}\prod_{j=1}^p\frac{\vartheta_1\left(t_j|-\frac{1}{\tau}\right)}{2\sin\left(\pi t_j\right)}\right)\\
&\hspace{0.3in}\sum_{\lambda\in\tilde{\Lambda}[R',R]}e^{i\left(\Delta\theta+\pi s\right)\cdot \lambda}\tilde{q}^{\frac{\lambda^2}{2}}.
\end{split}
\end{equation}
The first bracketed term above corresponds to the $U(1)^M$ symmetry, the second corresponds to the $-1$ eigenvalues and the third term in brackets is the contribution coming from complex conjugate eigenvalues. This partition function may finally be written in the open string channel as
\begin{equation} \label{eq4.11}
\begin{split}
Z^{\Delta\theta}_{R',R}=\frac{g_{R'}g_{R}}{\mathrm{Vol}\left(\tilde{\Lambda}[R',R]\right)}\sqrt{\mathrm{det}'\left(\mathbb{I}-R'^{T}\cdot R\right)}&\left(\frac{1}{\eta\left(\tau\right)^M}\right)\left(\frac{\eta\left(\tau\right)}{\eta\left(\frac{\tau}{2}\right)}\right)^{N-M-2p}\\
& \left(\prod_{j=1}^p\frac{i\eta(\tau)}{\vartheta_1\left(t_j,\tau|\tau\right)}\right)\sum_{Q\in\sigma}q^{\frac{Q^2}{2}},
\end{split}
\end{equation}
where $Q$ denotes the $U(1)^M$ charges. It takes the values,
\begin{equation} \label{eq4.12}
Q=\mu+\prod\left(\frac{\Delta\theta}{2\pi}+\frac{s}{2}\right),
\end{equation}
where $\mu \in \Lambda^{*}[R',R]$ and $\prod$ denotes the projection of vectors into the subspace $\Lambda[R',R]$. This also defines the set $\sigma$ in eq.\eqref{eq4.11}. The notation $\mathrm{det}'$ in eq.\eqref{eq4.11} means that the determinant is over the non-vanishing eigenvalues. Let us denote
\begin{equation} \label{eq4.13}
G[R',R]=\frac{g_{R'}g_{R}}{\mathrm{Vol}\left(\tilde{\Lambda}[R',R]\right)}\sqrt{\mathrm{det}'\left(\mathbb{I}-R'^{T}\cdot R\right)},
\end{equation}
the quantity $G[R',R]$ is conjectured to take values in $\mathbb{Z}\cup\sqrt{2}\mathbb{Z}$. For the cases where $G[R',R]\in\sqrt{2}\mathbb{Z}$, there is an unpaired Majorana fermion and hence the boundary condition are not compatible and are thus not considered. In the cases where $G[R',R]\in\mathbb{Z}$, it denotes the degeneracy of the ground states.
\subsection{Entanglement Spectra: Total and Symmetry resolved} \label{section4.2}
We now proceed to study the total and symmetry-resolved entanglement spectrum of the theory. We will first consider the boundary conditions preserving the same $U(1)^N$ chiral symmetry on either end. The entanglement spectrum, using eq.\eqref{eq4.8}, is given by
\begin{equation} \label{eq4.14}
\begin{split}
S_n=\frac{n+1}{12n}NW+\log\left(\mathrm{Vol}(\Lambda[R])\right)+&\frac{N}{1-n}\sum_{k=1}^{\infty}\log\left(\frac{\left(1-\tilde{q}^k\right)^n}{1-\tilde{q}^{\frac{k}{n}}}\right)\\
&+\frac{1}{1-n}\log\left(\frac{\sum_{\lambda\in\Lambda[R]}\tilde{q}^{\frac{\lambda^2}{2n}}e^{i\Delta\theta\cdot\lambda}}{\left(\tilde{q}^{\frac{\lambda^2}{2}}e^{i\Delta\theta\cdot\lambda}\right)^n}\right)
\end{split}
\end{equation}
\begin{equation} \label{eq4.15}
\begin{split}
S_1=N\frac{W}{6}+\log\left(\mathrm{Vol}(\Lambda[R])\right)&-N\sum_{k=1}^{\infty}\left(\log(1-\tilde{q}^k)-k\log\left(\tilde{q}\right)\frac{\tilde{q}^k}{1-\tilde{q}^k}\right)\\
&+\frac{1}{2}\log\left(\tilde{q}\right)\frac{\sum_{\lambda\in\Lambda[R]}\lambda^2\tilde{q}^{\frac{\lambda^2}{2}}e^{i\Delta\theta\cdot\lambda}}{\sum_{\lambda\in\Lambda[R]}\tilde{q}^{\frac{\lambda^2}{2}}e^{i\Delta\theta\cdot\lambda}}+\log\left(\sum_{\lambda\in\Lambda[R]}\tilde{q}^{\frac{\lambda^2}{2}}e^{i\Delta\theta\cdot\lambda}\right)
\end{split}
\end{equation}
As expected, the leading order contribution is proportional to the central charge N. The subleading term, i.e. $\log\left(\mathrm{Vol}(\Lambda[R])\right)$ term, is the Affleck-Ludwig boundary entropy term. The rest of the contributions vanish in the asymptotic limit, similar to the case of a single Dirac fermion. A notable difference from the single fermion case is the boundary entropy term present here. This term however vanishes when $\mathrm{Vol}(\Lambda[R])=1$, which holds when we consider the $U(1)$ symmetry of kind given by eq.\eqref{eq3.8}-\eqref{eq3.9} for each of the N fermions. Now let's proceed to study the symmetry-resolved entanglement spectrum. Utilising the diagonal representation of the partition function in eq.\eqref{eq4.8} in the $U(1)^N$ charge sectors, the symmetry resolved entanglement entropies are given by
\begin{equation} \label{eq4.16}
\begin{split}
S_n(Q)=&\frac{N}{1-n}\log\left(\frac{\eta\left(-\frac{1}{n\tau}\right)}{\eta\left(-\frac{1}{\tau}\right)^n}\right)\\
=&N\left(\frac{1+n}{n}\frac{W}{12}-\frac{1}{2}\log\left(\frac{W}{\pi}\right)+\frac{1}{2(1-n)}\log(n)+\frac{1}{1-n}\sum_{k=1}^{\infty}\log\left[\frac{\left(1-\tilde{q^k}\right)^n}{1-\tilde{q}^{k/n}}\right]\right).
\end{split}
\end{equation}
This is just the result of a single Dirac fermion obtained in eq.\eqref{eq3.17} with an overall factor of $N$. The symmetry resolved entanglement entropy $S_{1}(Q)$ is also given by $N$ times the expression in eq.\eqref{eq3.18}. It is interesting to note that while the entanglement spectrum depends upon the $U(1)^N$ chiral symmetry preserved even in the asymptotic limit, the symmetry-resolved entanglement doesn't.

We now consider the boundary conditions on either end such that they preserve a $U(1)^M$ subgroup of either $U(1)^N$ chiral symmetry. The entanglement spectrum in this case may be found using the mixed boundary condition partition function given by eq.\eqref{eq4.10}. The entanglement entropies are found to be
\begin{equation} \label{eq4.17}
\begin{split}
S_n&=N\frac{1+n}{12n}W+\log\left(G[R',R]\right)-\frac{M-N-2p}{2}\log(2)+\log\left(\mathrm{Vol}\left[\tilde{\Lambda}[R',R]\right]\right)\\
&+\frac{1}{1-n}\log\left(\frac{\sum_{\lambda\in\tilde{\Lambda}}e^{i\left(\Delta\theta+\pi s\right)\cdot\lambda}\tilde{q}^\frac{\lambda^2}{2n}}{\left(\sum_{\lambda\in\tilde{\Lambda}}e^{i\left(\Delta\theta+\pi s\right)\cdot\lambda}\tilde{q}^\frac{\lambda^2}{2}\right)^n}\right)+\frac{N-2p}{1-n}\sum_{k=1}^{\infty}\log\left(\frac{\left(1-\tilde{q}\right)^n}{1-\tilde{q}^{\frac{1}{2}}}\right)\\
&-\frac{N-M-2p}{1-n}\sum_{k=1}^{\infty}\log\left(\frac{\left(1-\tilde{q}^2\right)^n}{1-\tilde{q}^{\frac{2}{2}}}\right)+\frac{1}{1-n}\sum_{j=1}^{p/2}\sum_{k=1}^{\infty}\log\left(\frac{1-2\cos\left(2\pi t_j\right)\tilde{q}^{\frac{1}{n}}+\tilde{q}^{\frac{2}{n}}}{\left(1-2\cos\left(2\pi t_j\right)\tilde{q}+\tilde{q}^{2}\right)^n}\right).
\end{split}
\end{equation}
Taking the $n\to 1$ limit is straightforward, we obtain expressions similar to eq.\eqref{eq4.15}. We don't obtain any new behaviour in this limit as well, so for brevity we omit writing its expression. The leading order behaviour is again proportional to $N$. At subleading order, i.e. the $O(1)$, we have three contributions. The term $\log\left(G[R',R]\right)$ is due to the degeneracy of the $U(1)^M$ representations in the open string channel. The $\log\left(\mathrm{Vol}\left[\tilde{\Lambda}[R',R]\right]\right)$ is somewhat similar to the boundary entropy in the $U(1)^N$ case, however here it is not the only contribution to the boundary entropy. Finally, the third contribution $\frac{M-N-2p}{2}\log(2)$ is proportional to the number of $-1$ eigenvalues of $R'^{T}\cdot R$, which also corresponds to a remnant $Z_2$ symmetry, a similar term also appears in the case of mixed boundary conditions in the modular invariant Dirac fermion, see eq.\eqref{eq3.39}. These contributions suggest that the entanglement spectrum is sensitive to both the symmetries preserved and symmetries broken by the boundary conditions. The remaining terms in eq.\eqref{eq4.17} vanishes in the asymptotic limit. Let's now proceed to study the symmetry-resolved entanglement spectrum for the preserved $U(1)^M$ symmetry. We again use the fact that the partition function in the open string channel is diagonal in $U(1)^M$ charges, see eq.\eqref{eq4.11}-\eqref{eq4.12}. The symmetry resolved entanglement entropies are given by 
\begin{equation} \label{eq4.18}
\begin{split}
S_n(Q)&=N\frac{1+n}{12n}W+\log\left(G[R',R]\right)-\frac{M-N-2p}{2}\log(2)+\frac{M}{2(1-n)}\log(n)\\
&-\frac{M}{2}\log\left(\frac{W}{\pi}\right)+\frac{N-2p}{1-n}\sum_{k=1}^{\infty}\log\left(\frac{\left(1-\tilde{q}\right)^n}{1-\tilde{q}^{\frac{1}{2}}}\right)-\frac{N-M-2p}{1-n}\sum_{k=1}^{\infty}\log\left(\frac{\left(1-\tilde{q}^2\right)^n}{1-\tilde{q}^{\frac{2}{n}}}\right)\\
&\hspace{2.2in}+\frac{1}{1-n}\sum_{j=1}^{p/2}\sum_{k=1}^{\infty}\log\left(\frac{1-2\cos\left(2\pi t_j\right)\tilde{q}^{\frac{1}{n}}+\tilde{q}^{\frac{2}{n}}}{\left(1-2\cos\left(2\pi t_j\right)\tilde{q}+\tilde{q}^{2}\right)^n}\right).
\end{split}
\end{equation}
We see that similar to the $U(1)^N$ case, the leading order contributions are $\log\left(\frac{\ell}{\epsilon}\right)$, and $\log\log\left(\frac{\ell}{\epsilon}\right)$ with the former term being proportional to $N$ and latter being proportional to $M$. The double log term is characteristic of $U(1)$ symmetry resolved entanglement. At $O(1)$, we however see new terms appearing, $\log\left(G[R',R]\right)$, and $\frac{M-N-2p}{2}\log(2)$, both these terms were also present in the entanglement spectrum given by eq.\eqref{eq4.17}. Similar to the $U(1)^N$ case, the $\log\left(\mathrm{Vol}\left[\Lambda\right]\right)$ term is absent in eq.\eqref{eq4.18}. The symmetry-resolved spectrum continues to be independent of the charge sector to all order in this case as well.
\section{Conclusion} \label{section5}
In this work, we studied the total and symmetry-resolved R\'enyi entropies of some Fermionic CFTs for a single interval. The BCFT approach allows us to study all the universal contributions to both the entanglement spectra, given that the boundary conditions respect the symmetry under scrutiny. Here, we first considered the $2$d massless Dirac fermions and modular invariant Dirac fermions with the boundary conditions preserving either the vector $U(1)$ symmetry, denoted $U_{V}(1)$, or the axial $U(1)$ symmetry, denoted $U_{A}(1)$, but not both as they have an anomaly between them. In the massless Dirac fermion case, both the R\'enyi entropies were independent of the boundary conditions. The only difference between the two boundary conditions was that of the charge spectrum of the two currents. We also noticed that the boundary entropy term was absent in R\'enyi entropies. The symmetry-resolved R\'enyi entropies of the modular invariant Dirac fermion were found to match the massless Dirac fermion result for either boundary conditions. Here again, the charge spectra of the two currents differed. The R\'enyi entropies for the two boundary conditions differed in the boundary entropy and charge spectrum contribution in modular invariant theory. The results for $U_A(1)$ were also shown to match with the Dirac fermions. Finally, the entropies of the modular invariant Dirac fermions were shown to match the corresponding entropies of the compact boson at the duality radius.

We next considered $N$ massless Dirac fermions with the boundary conditions preserving $U(1)^N$ chiral symmetries, using the results of \cite{smith2020boundary}. We studied both the R\'enyi entropies by imposing different boundary conditions on either end of the interval such that a $U(1)^M$ subgroup of either $U(1)^N$ chiral symmetry was preserved, where $M\leq N$. The R\'enyi entropies were found to have contributions at $O(1)$ sensitive to the degeneracy of the $U(1)^M$ characters in the open string channel and another one sensitive to symmetry broken by the boundary conditions. We also found a boundary entropy-like term at the same order. This term was similar to the boundary entropy term in the case where both the boundary conditions preserved the same $U(1)^N$ chiral symmetry. The symmetry resolved R\'enyi entropies were shown to have the characteristic $\log\log\left(\frac{\ell}{\epsilon}\right)$ term for the $U(1)$ symmetry proportional to $M$. Finally, at $O(1)$ we found a contribution sensitive to the degeneracy of the $U(1)^M$ representation in the symmetry resolved spectrum as well.

The BCFT approach allowed us to observe some interesting contributions to both the entanglement spectra. It would be interesting to use this approach in theories possessing a higher symmetry group that have interesting representations in the boundary theory. Wess-Zumino-Witten models would be an ideal candidate, we mention however that the symmetry resolved entanglement has been studied for these models in \cite{calabrese2021symmetry}, however, the BCFT approach was not considered there. We believe minimal models would be another interesting candidate for a detailed study as well in this respect.   
\appendix
\section{Modular functions} \label{A}
In this appendix, we give the definition of Jacobi theta function and Dedekind's eta function. We also give their modular properties under the S transformation used in the main text.

The Jacobi theta functions and the Dedekind's eta function are given by the following expressions
\begin{align}
\vartheta_1(z|\tau)&=i\sum_{n=-\infty}^{\infty}(-1)^{n}q^{\frac{1}{2}\left(n-\frac{1}{2}\right)^2}e^{i2\pi \left(n-\frac{1}{2}\right)z}\\
\vartheta_2(z|\tau)&=\sum_{n=-\infty}^{\infty}q^{\frac{1}{2}\left(n-\frac{1}{2}\right)^2}e^{i2\pi \left(n-\frac{1}{2}\right)z}\\
\vartheta_3(z|\tau)&=\sum_{n=-\infty}^{\infty}q^{\frac{n^2}{2}}e^{i2\pi nz}\\
\vartheta_4(z|\tau)&=\sum_{n=-\infty}^{\infty}(-1)^nq^{\frac{n^2}{2}}e^{i2\pi nz}\\
\eta\left(\tau\right)&=q^{\frac{1}{24}}\prod_{n=1}^{\infty}\left(1-q^n\right),
\end{align}
where $q=e^{i2\pi\tau}$, $\tau$ being the modular parameter. The S transformation, that is $\tau\to -\frac{1}{\tau}$, properties of these functions may be determined using the Poisson re-summation formula, these identities are given below 
\begin{align}
\vartheta_1\left(z|-\frac{1}{\tau}\right)&=-i\sqrt{-i\tau}e^{i\pi\tau z^2}\vartheta_1\left(z\tau|\tau\right)\\
\vartheta_2\left(z|-\frac{1}{\tau}\right)&=\sqrt{-i\tau}e^{i\pi\tau z^2}\vartheta_4\left(z\tau|\tau\right)\\
\vartheta_3\left(z|-\frac{1}{\tau}\right)&=\sqrt{-i\tau}e^{i\pi\tau z^2}\vartheta_3\left(z\tau|\tau\right)\\
\vartheta_4\left(z|-\frac{1}{\tau}\right)&=\sqrt{-i\tau}e^{i\pi\tau z^2}\vartheta_2\left(z\tau|\tau\right)\\
\eta\left(-\frac{1}{\tau}\right)&=\sqrt{-i\tau}\eta(\tau),
\end{align}
\section{Boundary States for Dirac Fermion} \label{B}
In this appendix, we find the Cardy boundary states for massless Dirac fermions and modular invariant Dirac fermions with either boundary condition given by eq.\eqref{eq3.8}-\eqref{eq3.9}. The boundary states for the Neveu-Schwarz ($\mathrm{NS}$) sector for either boundary conditions are known, for completeness we give a derivation of these states in both the sectors here.

Let us first give the constructive Bosonisation of the fermions in either sector following \cite{von1998bosonization} on the complex plane. We define 
\begin{equation} \label{eqB.1}
\psi(z)=\hat{F}^{\dagger}z^{\hat{N}}\hat{\lambda}e^{-i\varphi^{\dagger}(z)}e^{i\varphi(z)},
\end{equation}
where $\hat{F}$ is the ladder operator, also called Klein factors, acting on the fermion number eigenstates as $\hat{F}\left|N,\bar{N}\right\rangle=\left|N-1,\bar{N}\right\rangle$. The operator $\hat{N}$ may be thought of as the fermion number operator and $\hat{\lambda}$ is the phase counting operator. The phase counting operator is needed due to the anti-commutation relation of the fermion operators and our convention of the fermion states. The boson operator $e^{i\varphi(z)}$ in eq.\eqref{eqB.1} conserves the fermion number, but creates excitations on the fixed fermion number ground state. In the Ramond sector, we pick our reference states in the holomorphic sector to be $\left|\sigma\right\rangle$. We also define fermion number operator $\hat{N}$ such that $\hat{N}\left|\sigma\right\rangle=0$ and $\hat{N}\left|\rho\right\rangle=-\left|\rho\right\rangle$ in this sector, it has the relation $\hat{N}=J_0-\frac{1}{2}$ with the $U(1)$ current. In the Neveu-Schwarz sector, we have $\hat{N}=J_0$ and we have only one choice for the reference ground state. The phase counting operator is defined as
\begin{align}
\hat{\lambda}\left|N,\bar{N}\right\rangle=&\left|N,\bar{N}\right\rangle \label{eqB.2}\\
\hat{\bar{\lambda}}\left|N,\bar{N}\right\rangle=&(-1)^{N}\left|N,\bar{N}\right\rangle, \label{eqB.3}
\end{align} 
where the states $\left|N,\bar{N}\right\rangle$ are the fermion number eigenstates satisfying $\hat{N}\left|N,\bar{N}\right\rangle=N\left|N,\bar{N}\right\rangle$, and $\hat{\bar{N}}\left|N,\bar{N}\right\rangle=\bar{N}\left|N,\bar{N}\right\rangle$ in either sector.

To construct the boundary states given in Section \ref{section3}, we impose the boundary conditions $\alpha$ at $|z|=1$ on the complex plane and consider the boundary theory on $|z|\geq 1$. The Cardy boundary states must satisfy the Cardy conditions \cite{cardy1989boundary} and the Cardy-Lewellen sewing conditions \cite{lewellen1992sewing, cardy1991bulk}. In the present case, the Cardy-Lewellen sewing conditions implies the clustering
\begin{equation} \label{eqB.4}
\lim_{|z|,|w|\to 1} \frac{\left\langle\mathcal{O}_p(z)\mathcal{O}_q(w)\right\rangle_{\alpha}}{\left\langle\mathcal{O}_p(z)\right\rangle_{\alpha} \left\langle\mathcal{O}_q(w)\right\rangle_{\alpha}}=1,
\end{equation}
where $\mathcal{O}_p(z)$ is a composite operator constructed from the fermion operators via some appropriate regularisation scheme. The clustering conditions in eq.\eqref{eqB.4}, using the boundary state, may be expressed as
\begin{equation} \label{eqB.5}
\lim_{|z|,|w|\to 1}\frac{\left\langle 0\right|\mathcal{O}_p(z)\mathcal{O}_q(w)\left|\alpha\right\rangle\left\langle 0\right.\left|\alpha\right\rangle}{\left\langle 0\right|\mathcal{O}_p(z)\left|\alpha\right\rangle \left\langle 0\right|\mathcal{O}_q(w)\left|\alpha\right\rangle}=1,
\end{equation} 
where $\left|\alpha\right\rangle$ is the Cardy boundary state in the extended theory. The composite operator $\mathcal{O}_p(z)$ may be thought of as a chiral vertex operator taking the $\hat{N}$ eigenstate $\left|p\right\rangle$ to the reference ground state $\left| 0\right\rangle$. The one-point correlation function of the composite operator $\mathcal{O}_p$ is non-vanishing only when it has the corresponding anti-holomorphic part with $U(1)$ charge index $\bar{p}=\tilde{\alpha}(p)$, where $\tilde{\alpha}$ is the automorphism induced on $\bar{\mathcal{H}}\to\mathcal{H}$ under the boundary condition $\alpha$. In other word $\bar{p}=\tilde{\alpha}(p)$ is conjugate to $p$ under $\alpha$. Therefore, we only consider the composite operators $\mathcal{O}_p(z)$ that possesses the anti-holomorphic part $\bar{p}$ conjugate to the holomorphic part $p$.
The Cardy state is expressed as a linear combination of the Ishibashi states, see eq.\eqref{eq3.11}-\eqref{eq3.12}. Let us write
\begin{equation} \label{eqB.6}
\left|\alpha\right\rangle=\sum_{p}a_{p}\left|\left|p,\bar{p}\right\rangle\right\rangle.
\end{equation}
Using the expression of Cardy states in eq.\eqref{eqB.6} and our assertion that there are composite operators that behave as the chiral vertex operators in the clustering limit, the clustering conditions in eq.\eqref{eqB.5} reduces to
\begin{equation} \label{eqB.7}
\frac{a_0a_{p+q}}{a_pa_q}\frac{\left\langle 0\right|\left(\hat{\bar{F}}\hat{\bar{\lambda}}\right)^{\bar{p}+\bar{q}}\left(\hat{F}\hat{\lambda}\right)^{p+q}\left|p+q,\bar{p}+\bar{q}\right\rangle}{\left\langle 0\right|\left(\hat{\bar{F}}\hat{\bar{\lambda}}\right)^{\bar{p}}\left(\hat{F}\hat{\lambda}\right)^{p}\left|p,\bar{p}\right\rangle\left\langle 0\right|\left(\hat{\bar{F}}\hat{\bar{\lambda}}\right)^{\bar{q}}\left(\hat{F}\hat{\lambda}\right)^{q}\left|q,\bar{q}\right\rangle}=1
\end{equation}
This condition allows us to relate $a_p$ with $a_0$. In the $\mathrm{NS}$ sector we obtain the following relations for either boundary conditions $\mathrm{A}$ or $\mathrm{V}$
\begin{equation} \label{eqB.8}
a_p=e^{i\theta p+i\pi\frac{ p^2}{2}}a_0,
\end{equation}
where $\theta$ appearing here is the same as in eq.\eqref{eq3.8}-\eqref{eq3.9}. In the $\mathrm{R}$ sector, the reference ground states are taken to be $\left|\sigma,\bar{\sigma}\right\rangle$, and $\left|\sigma,\bar{\rho}\right\rangle$ for $\mathrm{V}$, and $\mathrm{A}$ boundary conditions respectively. We accordingly define the number operator $\hat{\bar{N}}$, satisfying $\hat{\bar{N}}\left|\bar{\sigma}\right\rangle=0$, and $\hat{\bar{N}}\left|\bar{\rho}\right\rangle=0$ for $\mathrm{V}$, and $\mathrm{A}$ boundary conditions respectively. We obtain the following relations from eq.\eqref{eqB.7} for either boundary condition
\begin{equation}
\tilde{a}_p=e^{i\theta p+i\pi\frac{ p^2}{2}}\tilde{a}_0.
\end{equation}
This is the same as in the $\mathrm{NS}$ sector, however, we use different notations to emphasize that these coefficients are not necessarily the same in both sectors. The coefficients $a_0$ and $\tilde{a}_0$ are determined from the Cardy conditions and are given in Section \ref{section3}, it should also be noted that the coefficients $a_0$ and $\tilde{a}_0$ also depend on the boundary conditions as well.
\acknowledgments 
HG is supported by the Prime Minister’s Research Fellowship offered by the Ministry of
Education, Govt. of India. HG would also like to thank Urjit A. Yajnik for the helpful discussions, and comments on the manuscript. We would also like to thank the anonymous Referee for their insightful comments on the manuscript.

\bibliographystyle{unsrt}
\bibliography{Bibtex1.bib}



\end{document}